\documentclass[12pt,preprint]{aastex}



\def\asca       {{\em ASCA}\/}
\def\chandra    {{\em Chandra}\/}

\def\rosat      {{\em ROSAT}\/}
\def\sax        {{\em BeppoSAX}\/}
\def\kms{km s$^{-1}$}

\begin{document}

\title{{\em CHANDRA} OBSERVATION OF A2256 --- A CLUSTER AT THE EARLY STAGE OF MERGING}

\author{M.\ Sun, S.\ S.\ Murray, M.\ Markevitch \& A. Vikhlinin\altaffilmark{1}}
\affil{Harvard-Smithsonian Center for Astrophysics, 60 Garden St.,
Cambridge, MA 02138;\\ msun@cfa.harvard.edu}

\altaffiltext{1}{Also Space Research Institute, Moscow, Russia}

\shorttitle{\chandra\ observation of A2256}
\shortauthors{SUN ET AL.}

\begin{abstract}

We present \chandra\ observations of the rich cluster of galaxies
A2256. In addition to the known cool subcluster, a new structure (we
call it "shoulder" in this paper based on its morphology) was resolved
2$'$ east of the peak of the main cluster. It is shown as a localized 
feature embedded in the main cluster. Its position is roughly at the
center of a low-brightness radio relic. Spectral analysis shows that
the "shoulder" has high iron abundance $\sim$ 1 (after the decomposition).
The gas mass within
it is around 2$\times$10$^{11}$ M$_{\rm \odot}$. We suggest that this
structure is either another merging component or an internal structure
of the main cluster. The previously known subcluster has a low temperature ($\sim$ 4.5 keV)
and high iron abundance ($\sim$ 0.6) in the central 150 kpc. The main cluster
has the temperature of 7 - 8 keV and the iron abundance of 0.2 - 0.3 around
the center. The \chandra\ image shows a relatively sharp brightness
gradient at the south of the subcluster peak running south-south-east (SSE).
A temperature jump was found across the edge, with lower temperature
inside the subcluster. This phenomenon is qualitatively similar to the
"cold fronts" found in A2142 and A3667. While a simple interpretation is not
possible due to the projection, the edge indicates relative motion and
contact of the two gas clouds. The \chandra\ temperature map shows
only moderate temperature variations across
the cluster, but not as strong as those expected in a major merger. If the
"shoulder" is ignored, the temperature map resembles those simulations at
the early stage of merging while the subcluster approached the main cluster
from somewhere west. The observed temperature map and the edge-like feature
near the south of the subcluster, imply that the ongoing merger is still
at the early stage. The X-ray redshifts of several regions were measured.
The results are consistent with a single value and all agree with the
optical value. At least three member galaxies, including
a radio head-tail galaxy, were found to have corresponding X-ray emission
with X-ray luminosity from several times 10$^{40}$ to 0$^{41}$ ergs s$^{-1}$.
It is found that the observed characteristics (temperature and iron abundance
gradient) of the subcluster are similar to those of some poor clusters.
The absence of galaxies around the peak of the subcluster is proposed to be
the result of different falling velocity of the galaxies and the core gas.

\end{abstract}


\section{Introduction}

In the hierarchical scenarios of the structure formation of the universe,
cluster of galaxies are formed through subcluster
mergers. Both observations (e.g. Bliton et al. 1998; Owen et al. 1999; 
Markevitch et al. 2000) and simulations (e.g. Roettiger, Loken \& Burns 1997;
Roettiger, Stone \& Burns 1999; Bekki 1999) show that the merging dramatically
change the physical characteristics of the intracluster medium (ICM)
and galaxies, like the temperature, density and magnetic field strength
of ICM, star formation rate and radio emission of galaxies. However, many
processes related to the mergers and their effects are still vague.
Hence systematic analyses of merging clusters are very helpful to deepen our
understanding on this complex dynamical process.
Here we present recent \chandra\ observations and analysis on
the well-known merging cluster A2256.

A2256 is a rich cluster with strong X-ray emission (L$_{x} \sim$
10$^{45}$ ergs s$^{-1}$) at the redshift 0.058. \rosat\
observations (Briel et al. 1991) clearly revealed two peaks near the center,
separated by about $3.5'$. One of them is roughly at the geometric center
of the cluster, while
the other one is considered to be a merging subcluster. The previous
X-ray observations also showed that the subcluster has a lower temperature
than the main cluster (Briel et al. 1991; Miyaji et al. 1993; Markevitch
1996, M96 hereafter) and temperature generally decreases with the radius (M96;
Molendi, De Grandi \& Fusco-Femiano 2000, MDF00 hereafter). Briel \& Henry
(1994) reported two hot spots ($>$ 12 keV) in this cluster using \rosat\
data. However, they were not confirmed by \asca\
and \sax\ observations (M96; MDF00).
The velocity dispersion of the galaxies is very large ($\sim$ 1300 km/s; Fabricant,
Kent \& Kurtz 1989) and the velocity distribution
shows substructure (Roettiger, Burns \& Pinkney 1995).
The radio properties of A2256 are anomalous (Bridle \& Fomalont 1976;
Bridle et al. 1979; R$\ddot{\rm o}$ttgering et al. 1994, R94 hereafter). A
remarkable radio relic, with sharp edges and possible filamentary structure,
was found northwest of the two X-ray
peaks. At least four head-tail sources were found, including one with
an exceptional narrow, straight tail extending to at least about 0.7 Mpc.
R94 suggested that the halo was composed
of a few head-tail galaxies that were heavily distorted due to the
infalling subcluster.
Recent \sax\ observation (Fusco-Femiano et al. 2000) 
also showed the existence of nonthermal X-ray emission from the cluster
but only significant above 10 keV.

In this paper, we present the analysis of the recent \chandra\ observations
of A2256. \chandra\ observations and data reduction are in
$\S$2. The analysis is in $\S$3. $\S$4 is the discussion and $\S$5 is the summary.
Throughout this paper, we assume H$_{0}$ = 70 km s$^{-1}$ Mpc$^{-1}$ and
q$_{0}$ = 0.5. These cosmological parameters correspond to the linear scale
of 1.08 kpc/arcsec at the cluster redshift.
 
\section{Observations and data reduction}

A2256 was observed by \chandra\ three times with Advanced CCD Imaging
Spectrometer (ACIS). The information of the observations are listed
in Table 1. For each observation, we excluded the known bad columns, hot
pixels, chip node boundaries and events with \asca\ grades 1, 5 and 7,
as well as bad aspect intervals. The proper gain map file for that period
was used. Particle background flare periods were excluded based on the
X-ray light curves from the outer
parts of the field. The back-illuminated (BI) chips and front-illuminated
(FI) chips were filtered separately since they have different background
flare levels (see Markevitch et al. 2000 for details). The chips used
for data analysis are I0 - I3 chips of the ACIS-I observation and S2 - S4 chips
of the ACIS-S observation. The streak events in S4 chip were removed by the
"destreak" tool in CIAO. The short observation
1521 was dominated by background flares (over 80\% of its exposure time) and
the statistics are quite poor so we do not include that data in our
analysis.

To subtract the correct background, the blank field background dataset
relevant for the period of the observations was used (Markevitch 2001).
Since the background may change slightly with time, we also checked the
background rate using the data at high energy (10-12 keV in the
front-illuminated - FI chips and 9-12 keV in the back-illuminated -
BI chips). It was found that the background normalization should be
increased by 8\% in FI chips, consistent with the expected long-term
background trend. Note that although the correction should apply only
to the particle component of the total background (and not the cosmic
x-ray background or CXB), that component is dominant at energy greater
than 2 - 3 keV which is most important for the gas temperature fitting.
Small rescaling of the CXB component generally has only tiny effect on
the spectral fitting. Thus we apply this
correction across the energy range for simplicity. The systematic
uncertainty of the background normalization ($\pm$10\%) that encompasses
this correction was also included in all the confidence intervals 
reported below.

Shortly after the launch of \chandra, the ACIS FI
chips suffered increasing "charge transfer inefficiency"
(CTI) problem. The effect is seen as a decreasing quantum efficiency (QE)
with increasing distance from the readout node. Here the results of the
calibration observations of G21.5-0.9 were adopted to make CTI corrections
(Vikhlinin 2000). We also used an additional position-independent
correction factor of 0.93 for the ACIS-I quantum efficiency below
1.8 keV to account for the difference between ACIS-S3 and ACIS-I
(Vikhlinin 2000). For hot clusters, ignoring it results in spuriously
high temperature and the dependence of fitting results on the adopted
low energy cut (Markevitch \& Vikhlinin 2001).
CIAO(1.1.5), FTOOLS(5.0), XSPEC(10.0) and some our own software
were used to do the data reduction. The exposure maps were generated
using our own software (equivalent to the similar one in CIAO).
Since vignetting is dependent on the energy, narrow bands were
used and weighting spectra were also applied in producing
the exposure maps (the weighting spectra differ from chip to chip
based on the fitting result of the integrated spectrum of each chip).
To produce response files of the spectra in the interested regions
(response matrices - RMF and auxiliary response files - ARF), the
tools calcrmf and calcarf by Vikhlinin were used. The ARF was calculated
by weighting the mirror effective area in the region with the observed
cluster brightness distribution in 0.5 - 2.0 keV band. The RMF was
calculated by weighting the standard set of matrices within the region
by the observed cluster brightness distribution.
All the errors in this paper are 90\% confidence interval.

\section{Analysis}

\subsection{Point sources}

Over 30 point sources were detected in two observations using
the CIAO wavelet detection tool.
A detailed analysis on those point sources is beyond the scope of
the paper. In this section, we will mainly discuss the point sources
coincident with the member galaxies. Three member galaxies (all
ellipticals) were found to have corresponding X-ray sources in both
observations. They are shown in Fig. 1. One is the elliptical galaxy E
(Fig. 2) which is also a radio head-tail galaxy (A in R94). Only about 50
counts from it in total were collected during the two observations.
The statistics of the data do not allow us to constrain the temperature
or the photon index well even if its spectrum is that simple.
If we simply assumed a 1 keV thermal plasma with normal solar abundance,
the derived luminosity is about 8$\times$10$^{40}$ ergs s$^{-1}$ (0.5 - 10 keV).
If a powerlaw with photon index 1.7 is assumed, the derived luminosity is about 
3$\times$10$^{41}$ ergs s$^{-1}$ (0.5 - 10 keV).
The other two sources correspond to the NW core of the double galaxy NGC
6331 (C) and galaxy D respectively (Fig. 2).
Only about 20 counts in total were collected from each source in the two
observations. Their luminosity are estimated to be around 3$\times$10$^{40}$
ergs s$^{-1}$ for the 1 keV thermal plasma assumption, or 10$^{41}$
ergs s$^{-1}$ for the powerlaw (photon index 1.7) assumption (0.5 - 10 keV).
The X-ray emission may come from the low luminosity active galactic nuclei
(LLAGN) or thermal halos.

We also tried to explore the nature of the brightest two point
sources in the field (Fig. 1; \#1: 7.5 c/ks - I3; \#2: 5.4 c/ks
- I0 and 8.4 c/ks - S3). \#1 is also marginally seen in both \rosat\ PSPC
and HRI images. 
\#1 has a very faint optical counterpart in DSS II while \#2 has no counterpart.
We extracted and fitted their spectra with a simple powerlaw model.
If the absorption is fixed at the galactic value, the photon index for
\#1 and \#2 are 1.4$^{+0.5}_{-0.4}$ and 1.3$\pm$0.4 respectively.
Thus, they might be background AGNs.
Based on these best-fit, their 0.5 - 10 keV unabsorbed flux are
10$^{-13}$ ergs s$^{-1}$ cm$^{-2}$ and 7$\times$10$^{-14}$ ergs s$^{-1}$
cm$^{-2}$ respectively. 

We also compared the observed source number with the predicted by
the Log N - Log S
relation derived in \chandra\ deep fields (e.g. Giacconi et al. 2001)
and found no significant difference.

\subsection{Structure of diffuse emission}

The 0.5 - 7 keV combined ACIS-I/S image is shown in Fig. 1 overlaid
on a DSS II image. The background was subtracted and the image was
divided by the exposure map. All the point sources were excluded.
The two previously resolved peaks at the center (P$_{1}$ and P$_{2}$
hereafter) by \rosat\ (Briel et al. 1991) are prominent and both show
internal structures (Fig. 2). It is noticed that the central part of
the subcluster (the western peak) is elongated along the east-west direction,
while the central part of the main cluster (the eastern peak) is elongated
along the north-south direction (Fig. 1 and 2).
These two peaks are separated by about $3'$ and P$_{2}$ is somewhat brighter. 
In DSS II image, there is no galaxy concentration around P$_{2}$.
Radio head-tail galaxies E, F and G (Fig. 2) and double galaxy
NGC 6331 (C) all have offsets to P$_{2}$ from $1'$ to
$2'$. The surface brightness peak of the main cluster is located
0.5$'$ north of a big elliptical galaxy (B), while galaxy A,
the galaxy with the most extended optical halo in A2256, is about 50 kpc east
of the peak.
Besides P$_{1}$ and P$_{2}$, a new structure was found significantly
to the east of P$_{1}$ in \chandra\ observations (Fig. 1 and
2). In this paper we call it P$_{3}$ or "shoulder" from its morphology.
This structure is shown as a small clump extended
from P$_{1}$ but with a clear local maximum and seems to be embedded in the
main cluster. The \rosat\ images were re-checked and this structure
was also found in both HRI and PSPC images though it is not very significant.
It was also detected in the wavelet analysis on PSPC data by Slezak,
Durret \& Gerbal (1994). This structure has interesting spectral
characteristics and will be discussed later.

We also zoom-in the central part of A2256 to see any substructures around
the center (Fig. 2). The wavelet decomposition tool (Vikhlinin, Forman \& Jones
1997) was applied to the image and the reconstructed image (the right one
in Fig. 2) is quite similar to the simply smoothed one (the left one in
Fig. 2). Both images show complex structures around the center.
P$_{1}$ and P$_{2}$ also show internal structures, which may also be
a support for some kind of ongoing dynamical process in this cluster.
However, the current data do not allow us to constrain the spectral
difference on such small scales.
Besides the "shoulder" and the substructures within P$_{1}$ and P$_{2}$,
there are still several interesting significant facts.
First, there are no X-ray enhancements around the central biggest ellipticals
A, B and C (the 3 $\sigma$ upper limits are all about 2$\times$10$^{41}$
ergs s$^{-1}$ at 0.5 - 10 keV, assuming 1 keV thermal spectrum and 0.5$'$
aperture size.). Second, the surface brightness
gradient at the south of P$_{2}$ is sharper than those in other directions
(more prominent in the unsmoothed image; also pointed out by Briel \& Henry, 1991). 
Third, there is a protrusion which extends from the center of P$_{2}$ to the
southwest.
As shown in Fig. 6, after removing the main cluster, the southern
parts become even sharper. In $\S$3.6, we will show that it may be
something like a cold front.

\subsection{Temperature map}

In view of the complex central structure mentioned in $\S$3.2
and the fact that A2256 is very likely a cluster in merging,
the knowledge of the temperature
distribution can reveal us the nature of the structures 
and the merging. The resolution of the temperature map one can
achieve for A2256 with \chandra\ is limited by statistics and
the PSF effect can be ignored completely.
We used the following method to obtain the temperature map.
First, the field was divided into 30 regions (as shown in Fig. 3)
with similar number of counts, then for each region we fitted the spectrum
to get the temperature (for region \# 11 - 30, we fitted the spectra
from two observations simultaneously). In view of the uncertainty	
of the ACIS low energy response, only 0.9-9 keV data were used and the
absorption was fixed to the galactic value 4.1$\times$10$^{20}$ cm$^{-2}$.
The best-fit temperatures are not sensitive to any small excess in
absorption.
The redshift was fixed at 0.058 and the abundance was fixed at 0.3 solar
(we used the solar abundance table by Anders \& Grevesse 1989), which is
the average value of previous observations (e.g. M96; MDF00).
The MEKAL code in XSPEC was used.
Second, we shifted the regions half-size along the parallel and perpendicular
directions to the orientation of the first set of regions to get two additional
similar temperature maps. The final one was obtained by averaging the above three.
Then the final temperature map was adaptively smoothed a little bit
for a better presentation
(Fig. 3). Two checks were performed. The first is the integrated
temperature of the central $8.3'$ square (ACIS-S3 chip FOV, see Fig. 4).
The result is 6.7$^{+0.2}_{-0.2}$ keV, consistent with
the central temperature ($\sim$ 7 keV) reported by previous observations
(e.g. Miyaji et al. 1993; Markevitch \& Vikhlinin 1997b; White 2000; MDF00).
The second
is that we did spectral fits at the same regions as those in \sax\ paper
(MDF00). Our results generally agree with theirs within 0.5 keV,
except in their region 2NW, where \sax\ spectrum may
suffer the contamination from the nearby high temperature
regions due to the large point spread function (PSF) of \sax\ 
($105''$ HPD for MECS at 1.5 keV).

On the 5 arcmin scale, the ACIS temperature map is in agreement with that of
\asca\ (M96). It shows moderate temperature variations across the cluster
but not as strong as expected in a major merger (e.g. Roettiger et al. 1997
and other simulations). Combining this temperature map with the image, no
shock was found in the field (a feature that is likely a "cold front" will
be discussed in $\S$3.6). The subcluster has lower temperature ($\sim$ 4 keV
at the coolest region) than the main cluster
and the shape of the low temperature regions strongly suggests that
the subcluster entered the main cluster from somewhere west. This temperature
map, if we omit the "shoulder", resembles those at the early stages of
merging in simulations (e.g. Roettiger et. al 1997; Takizawa 1999; Takizawa
2000). It is also interesting that the apparent coolest part of the
subcluster is about $2'$ west of its surface brightness peak.
The main cluster appears largely undisturbed, as concluded by
Markevitch \& Vikhlinin (1997a) from the \asca\ results.
The central region of the main cluster is generally cooler than its
outskirts covered by our FOV. This is more likely due to the projection of
the cooler subcluster and possibly the "shoulder", rather than the genuine
temperature gradient. There is a hot region ($\sim$ 9 keV; around regions \#5
and \#16 in Fig. 3) at the north of the main cluster. We notice that it is
just in positional coincidence with the eastern lobe of the radio relic (Fig. 1).
This might imply some kind of physical relation between the merging events
and the radio relics. It is also noticed
that there is a hot region near the south edge of the subcluster. In
$\S$3.6, we will show it might be related to a cold front.
The two hot spots reported by Briel \& Henry (1994) are located in
the FOV of ACIS-I (Fig. 3). Our measurements (7.9$^{+1.3}_{-1.0}$ keV for
their NE spot and 5.8$^{+1.5}_{-1.2}$ keV for the SW one) do not confirm the
existence of these hot spots, consistently with the earlier \asca\ 
and \sax\ conclusions (Markevitch \& Vikhlinin 1997a; MDF00).

We also checked the existence of any nonthermal component in the spectra,
especially at the regions of the radio relic. No clear evidence for
such component in ACIS spectra was found. A simple cooling flow component was
also added to the spectral model but that never changed $\chi^{2}$
significantly and the obtained mass accretion rates were always very small.

\subsection{Iron abundance \& red shift}

The integrated abundance for the central $8.3'$ square
(ACIS-S3 chip FOV) was obtained by simultaneously fitting the spectra
of two observations. It is 0.34$^{+0.06}_{-0.05}$, which is a little higher than
the previous results for the whole cluster, 0.27$\pm$0.06 by M96
(we converted their value for the abundance table used here)
and 0.25$\pm$0.03 by MDF00 (we converted their 68\%
confidence error into 90\%). This is consistent well with
the recent finding that the abundance generally falls with radius (MDF00) and their results in the inner radial bins.
Here we are more interested in the distribution of iron. The statistics
of the data do not permit a detailed analysis for the whole field. However,
in checking spectra from the regions that we used for the temperature map,
at least two with significantly high iron abundance areas were found.
The results are shown in Fig. 4. 
The four solid line regions, from a to d, represent P$_{1}$,
P$_{2}$, P$_{3}$ and a southern part of the main cluster
(relatively far from the subcluster) respectively.
The results from ACIS-I and ACIS-S agree well, so we performed
simultaneous fits for each region. As shown in the figure 4, 
it is clear that P$_{2}$ and P$_{3}$ have more iron than P$_{1}$ and
the southern main cluster region. The joint abundance-temperature 90\%
confidence regions for a - d are shown in Fig. 5, which again reinforces
that a and d are different from b and c.
Markevitch \& Vikhlinin (1997b) also obtained a higher iron abundance
at the subcluster than that of the main cluster using \asca\ data 
but the difference was not significant, which may due to the wide
PSF of \asca.
If we further consider the possible contamination of emission among the three
structures due to projection, even higher abundance contrasts
would be expected. The iron abundance in other regions
(e.g. the outer parts of the subcluster and the north part of the main cluster)
are all around the average ($\sim$ 0.3)
but poorly constrained. Current data can not allow us to determine
whether the high iron abundance is only localized in the "shoulder" or
also present in its immediate surrounding.
No iron abundance enhancement was found to be around galaxies C
and E.

We also measured the redshifts in several regions from the iron K$\alpha$
line blend. The regions we chose among those in Fig. 4 are b (the subcluster),
c (the "shoulder"), and the ACIS-S3 FOV excluding b, c and their $1'$
surrounding (the main cluster). The results are 0.056$^{+0.008}_{-0.007}$,
0.064$^{+0.007}_{-0.011}$, and 0.064$\pm$0.010 respectively. All the values
are consistent with a single redshift, suggesting that the main cluster,
the subcluster and the "shoulder" are possible to be associated. Using the
log N - log S relation for clusters (e.g. Kitayama, Sasaki \& Suto 1998;
De Grandi et al. 1999), and considering the observed maximal difference of
redshifts ($\sim$ 0.03), the possibility that they are a chance superposition
is less than 10$^{-4}$. Therefore we conclude that they are very likely
to be associated and interacting as also suggested by the temperature map
and the possible cold front ($\S$3.6).

The current data and calibration status do not allow us to constrain the spatial distribution of
other elements, like silicon. The integrated silicon abundance in
the ACIS-S3 FOV is 0.50$\pm$0.23 while the \asca\ result is
0.97$\pm$0.39 for the whole cluster (Fukazawa et al. 1998).

\subsection{Decomposition of the structures}

\chandra\ observations reveal complex structures around the center of
A2256. We should realize that due to the projection the observed spectra,
especially those around the center, have at least two components entangled
--- the main cluster and the subcluster.
It would be very helpful for our understanding if we can separate them.
However, this is actually very hard since the merging clusters are not
in hydrostatic equilibrium and we do not yet know their relative geometry.
Here, we just present a very simple way to isolate the subcluster
and decompose the spectral components.
Briefly speaking, we tried to use a $\beta$-model to represent the main
cluster (not the whole cluster) and examine the residuals. Then we can
use those information to separate the spectral components in observed spectra.
The \rosat\ image (background-subtracted and exposure-corrected)
shows that the surface brightness is rather spherical in the outer
parts, especially after we exclude the subcluster region using
a circle with some radius ($\sim$ 7$'$) centered at the core of the
subcluster. Thus we can simply measure the surface brightness beyond that
subcluster region to make a $\beta$-model fit to the main cluster.

The outer contours of the \rosat\ PSPC images were used to
find the geometric center of the main cluster, which is about $1'$ south
of the apparent peak of the main cluster. It is shown in Fig. 6.
Then a $7'$ radius circle centered at the peak of the subcluster was used
to represent the subcluster and a $2'$ radius circle at P$_{3}$ was used
to represent the emission from the "shoulder".
Only radial surface-brightness measurement outside of these circles
was made. All the point sources were excluded. ACIS-I/S
and PSPC data were used. The derived core radius is 3.8$\pm$0.3
arcmin (or 0.25$\pm$0.02 Mpc), and $\beta$ is 0.58$\pm$0.04.
The residual after removing the best-fit $\beta$-model of the main cluster
is shown in Fig. 6. The "shoulder" is more prominent and the morphology
of the subcluster is distorted, which is natural for a falling cluster.
Moreover, the southern surface brightness has become sharper and points
SSE, appearing more edge-like.

Then we can try to disentangle the spectral components in the observed
spectra. From the temperature map and the observed surface density
discontinuity ($\S$3.6), it is known that the subcluster enters the
main cluster from the west. Thus we do not expect much projection effect
from the subcluster on P$_{1}$. Here we only want to investigate P$_{2}$
and P$_{3}$ (the two regions in Fig. 2). To do that, we first need to
assume the temperature of the contaminating gas from the main cluster
at P$_{2}$ and P$_{3}$. Three value, 7, 8 and 9 keV, were assumed.
The abundance of the main cluster gas was fixed at 0.3.
The normalization of the contaminating gas is obtained from the
best-fit $\beta$-model.
The results are shown in Table 2. As expected, we obtained larger contrasts on
temperature and abundance between the main cluster and the others.
The abundance of the "shoulder" is still high after the decomposition.
From the best-fit emission measure, assuming a constant density sphere,
its gas mass is around
2$\times$10$^{11}$ M$_{\rm \odot}$ if its dimension along the line of sight
is not very different from others.

\subsection{The SSE edge --- a moving cold front?}

Fig. 6 shows a clear edge-like structure running along the SSE direction.
We measured the surface brightness profile in the linear regions
parallel to the edge in the co-added ACIS image (not the residual).
The temperature at each side of the edge are also obtained in the regions
shown in Fig. 6. The results are shown in Fig. 7.
At the edge, there is a break in the surface brightness profile
that indicates a density discontinuity
and a temperature jump from about 4.5 keV to about 8.5 keV.
This feature is quite similar to the cold fronts found in A2142 by
Markevitch et al. (2000) and A3667 by Vikhlinin et al. (2001a).
In A2142 and A3667, these features delineate the contact surfaces of the
cold dense clouds and the surrounding hotter gas, through which they move.
In the case of A2256, the geometry is more complicated and there is
projection of multiple structures. Therefore, it is not possible to 
derive the real density and pressure contrast across the edge.
The flat shape of the edge
may imply that it is only part of the whole moving edge (the other
parts are smoothed out by projection if the moving front is tilted
from the plane of sky), or the moving front viewed by an angle with
the line of sight. The protrusion mentioned in $\S$3.2 might be
related to the striping gas of the subcluster or its wake.

This discovery, adds another candidate to the list of merging clusters
which have cold fronts --- A2142, A3667 and RXJ1720.1+2638 (Mazzotta et al. 2001).
More cold fronts could be expected when more \chandra\ data are available.
Systematic studies of such phenomena will enrich our knowledge on
merging of the clusters.

\section{Discussion}

\subsection{The nature of the "shoulder"}

\chandra\ observations reveal a new structure - the "shoulder" - near the
center, a localized feature with size about $3'$ and gas mass about
2$\times$10$^{11}$ M$_{\rm \odot}$ (assuming a constant density sphere and its
dimension along the line of sight is similar as others).
Its spectrum implies that iron is enriched
in this structure ($\sim$ 1 after the decomposition).
Two galaxies (D and H in Fig. 2) are apparently located in this
structure though they are not necessarily associated with it.
Bridle \& Fomalont (1976) found a $10'$ low-brightness radio relic
roughly centered at galaxy D (also the radio source D in the
same paper) besides the bright NW halo (Fig. 1). The "shoulder"
is within that diffuse radio relic.

Although the roughly similar temperature and abundance between this
structure and the center of the subcluster may suggest that it
is a remaining of the subcluster spread along the path of its
infall, its location (east of P$_{1}$) is inconsistent with
the infalling direction (somewhere from the west) of the subcluster
suggested by the temperature map, the possible cold front and the general
direction of the radio head-tail sources (E, F and G in Fig. 2).

Since its X-ray redshift suggests it is a feature inside A2256 ($\S$3.4),
only two choices are left: either a new merging
component or an internal structure of the main cluster.
Another merging component seems to be a feasible explanation.
In fact, if we ignore the subcluster, the \chandra\ image resembles
simulated images for a merger between a massive cluster
and a much less massive one at about 0.3 Gyr after core crossing
(e.g. Takizawa 1999; Takizawa 2000). The fact that there
are two big elliptical galaxies rather than one dominant galaxy
around the center also supports the idea that the main cluster
has not yet relaxed well. Hence, we suggest that A2256 may be
a system with three clusters (or two clusters and one group) in merging.
About 0.3 Gyr before
the ongoing merger, there was a merger between the massive cluster
and a less massive system. The less massive system might be
a galaxy group that did not disturb the potential well of
the massive cluster much. The "shoulder" might be the relic of
its core and galaxy B might be once the dominant galaxy of that
less massive system since it is 2 times less luminous than A. 
Gonz\'alez-Casado, Mamon \& Salvador-Sol\'e (1994) found such kind of
the relic of the core can survive in at least one cluster crossing.
The observed iron abundance difference between the "shoulder" and
P$_{1}$ might suggest it was an off-center merger.
The less massive system may have fallen
into the massive cluster somewhere from the west as implied by the
simulations and the relatively sharper surface brightness at the east
than other directions. Under such scenario, the low-brightness 
radio relic found by Bridle \& Fomalont (1976) would be related
to this earlier merger event and the much brighter NW one may be related to
the ongoing merger. The fading of the radio haloes after the merger
was discussed by Tribble (1993). The typical aging time-scale is
at the order of 10$^{8}$ years, consistent with our picture here.

Another possibility is that the "shoulder" is an internal structure
of the main cluster as the two local dips of the potential around the
D galaxies in Coma (Vikhlinin et al. 2001b).
The gas mass of the "shoulder" is about 2$\times$10$^{11}$ M$_{\rm \odot}$,
while cD galaxies in clusters of galaxies often have gas content
at the level of 10$^{11}$ - 10$^{12}$ M$_{\rm \odot}$ (Trimble 2000).
The galaxy A, which resembles the cD most in A2256, is quite near the
"shoulder". Before the ongoing merger, 
the galaxy A may be at the center of the "shoulder", which could be the
halo of hot gas. When the ongoing merger began, the gases (the "shoulder")
might lag behind the galaxy A similar to the case discussed in $\S$4.2.
The iron mass excess within it is
estimated to be about 10$^{8}$ M$_{\rm \odot}$, while the stellar
mass of A is estimated to be about 7$\times$10$^{11}$ M$_{\rm \odot}$ assuming
M/L = 7(M/L)$_{\rm \odot}$ and L$_{\rm B}$ = 10$^{11}$ M$_{\rm \odot}$.
Thus, an injection of iron about 0.03\% of the stellar mass of A is
needed to explain the iron excess, which is still possible in the galactic-wind
models (e.g. Arimoto \& Yoshii 1987).

\subsection{The nature of the subcluster and the merger}

Fabian \& Daines (1991) suggested that the subcluster had a cooling flow
based on \rosat\ observations. In $\S$3.5 it is shown that the real
temperature of the subcluster is about 4.5 keV around the center.
The estimated gas density at the central 80 kpc of
the subcluster is about 3.6$\times$10$^{-3}$cm$^{-3}$ from the spectral
fitting. 
The cooling time scale is 85 Gyr (n/10$^{-3}$ cm$^{-3})^{-1}$
(T/8.6 keV)$^{1/2}$ , where n and T are the density and
temperature of the gas. Our revised cooling time at the center of the
subcluster is 14 - 20 Gyr. Though the estimated cooling
time cannot completely rule out the existence of the cooling flow, neither
the image nor spectra require any present cooling flow around the center of
the subcluster.

$\S$3.5 shows that the central 150 kpc of the subcluster has
high iron abundance ($\sim$ 0.6), which decreases
to about 0.3 at the outer parts. Briel \& Henry (1991) divided the
galaxies in A2256 into two groups using a simultaneous fit of
two Gaussians to the whole velocity data, though the two redshift distribution
components are not spatially separated.
Based on their fit, they suggested that the
subcluster was a poor cluster. Indeed, the observed X-ray characteristics of this
subcluster resemble those of some poor clusters (Virgo by Matsumoto et al. 1996;
Centaurus by Fukazawa et al. 1994; AWM 7 by Xu et al. 1997 and
Ezawa et al. 1997). All of them have comparably
low (or lower) temperature with the subcluster in A2256 and similar
abundance gradients. However, all these poor clusters have central cooling flows
while as discussed above, currently there is no cooling flow in the subcluster.
One possibility is that the cooling flow of the subcluster has been destroyed
by the early interaction during the merger while the abundance gradient was kept.

It is interesting that there is no galaxy
concentration near the surface brightness peak of the subcluster.
A natural explanation would involve the different
falling velocity of the galaxies (here they may be C, E, and F) and the
gas. The gas may lag behind the galaxies due to the drag
by the main cluster gas, which is implied by the highly distorted shape
of the subcluster gas and the apparent contact discontinuity.
We investigate whether this
picture is applicable to galaxy C, the best candidate of the cD of
the subcluster according to its halo size. Though both galaxies and
gas may undergo spiral-in, deceleration or acceleration, the time from when
they began to separate could be simply
expressed as 0.1 Mpc sin$^{-1}{\rm \theta}$ / (V$_{C}$ - V$_{G}$), where
${\rm \theta}$ is the angle with the line of sight, V$_{C}$ and V$_{G}$
are the velocity of the galaxy C and gas respectively. For a time scale
of 0.5 Gyr, a velocity difference about 200 sin$^{-1}{\rm \theta}$ \kms
can explain the observed displacement. Since large velocity dispersion
($\sim$ 1300 km s$^{-1}$) was found in this cluster, a velocity difference
(V$_{C}$ - V$_{G}$) around several hundred km s$^{-1}$ is quite
possible. 
 
The temperature map and the observed surface density discontinuity
imply that the subcluster entered the main cluster from somewhere west.
However, current data do not allow us to make a definite conclusion
to the moving direction of the subcluster. It has been suggested that
mergers may amplify the magnetic field and be responsible for the
strong radio haloes and relics (Tribble 1993; Roettiger et al. 1999). 
The NW bright radio relic in A2256 appears to be located
near the moving path of the subcluster, thus it is likely associated
with the ongoing merger event. Finally, the temperature map excludes
any significant disturbance of the cluster, indicating that the merger
with the subcluster is at very early stage.

\section{Summary}

\chandra\ observations of the merging cluster A2256 confirm some of
our previous understanding on it, but also reveal some new phenomena.
Our results are summarized below:

1) A new structure ("shoulder") was found $2'$ east of the peak of the main
cluster. It is shown as an extension from the main cluster peak
but with local maximum. Several galaxies (A, D and H in Fig. 2) are near or
apparently within it, but the physical association is hard to be established.
The position is also roughly at the center of a low-brightness radio relic.
This feature is also characterized by a high iron abundance
($\sim$ 1 after decomposition). Its temperature is comparable to or somewhat
lower than that of the surrounding. The gas mass within this structure
is about 2$\times$10$^{11}$ M$_{\rm \odot}$.
We suggest that it is either a relic of a prior merger or an internal
structure of the main cluster.

2) The subcluster was found to have a higher abundance ($\sim$ 0.6),
but a lower temperature (4.5 - 5 keV) than the main cluster (0.2 - 0.3 and
$\sim$ 7 keV respectively). Its characteristics resemble some of the poor
clusters.
The morphology of the subcluster is distorted, which could be due to the
infall. The central part of the subcluster, as well as the main cluster,
show some internal structures.

3) The brightness profile across the southern edge of the subcluster
indicates a density discontinuity and the temperature jump from about 4.5 keV
inside the dense gas to about 8.5 keV outside. This probably indicates
a contact the subcluster and the main cluster gases and their relative
motion, somewhat similar to A2142 and A3667 (Markevitch et al. 2000;
Vikhlinin et al. 2001a), although the geometry is more complex. Due to the
projection, the real moving direction of the whole edge is not easy to be
determined and the jumps of density and pressure are hard to be constrained.

4) The temperature map shows moderate temperature variation across
the cluster, but not as strong as expected in a major merger.
The temperature map implies that the subcluster entered the main cluster
from somewhere west and the merger is still at the early stage.
The main cluster appears not yet largely disturbed by the merger with
the subcluster. The two hot spots reported by Briel \& Henry (1994)
are not confirmed.

5) No X-ray enhancement is detected around the two brightest galaxies
(A and B in Fig. 2; the 3 $\sigma$ upper limits are about 2$\times$10$^{41}$
ergs s$^{-1}$ at 0.5 - 10 keV) around the center of the main cluster.
Over 30 point sources were detected in the observations, but only 1/3
of them have optical counterparts (usually faint) in the DSS II image.
No significant excess of point sources was found in the field.
Three member galaxies near the center, including a radio head-tail source
and a dumbbell-like big elliptical, were found to have corresponding
X-ray point-like emission. Their X-ray luminosity were estimated to be
several times 10$^{40}$ - 10$^{41}$ ergs s$^{-1}$.

\acknowledgments

The results presented here are made possible by the successful effort of the
entire \emph{Chandra} team to build, launch, and operate the observatory. We
are grateful to the referee for the valuable comments to improve the manuscript.
We acknowledge helpful discussions with W.\ Forman, C. Jones, D. M. Neumann,
T. Clarke and F. Durret. This study was supported by NASA contract NAS8-38248.

\clearpage

\begin{figure}
\vspace{-24mm}
\hspace{-1.9cm}
  \includegraphics[angle=270]{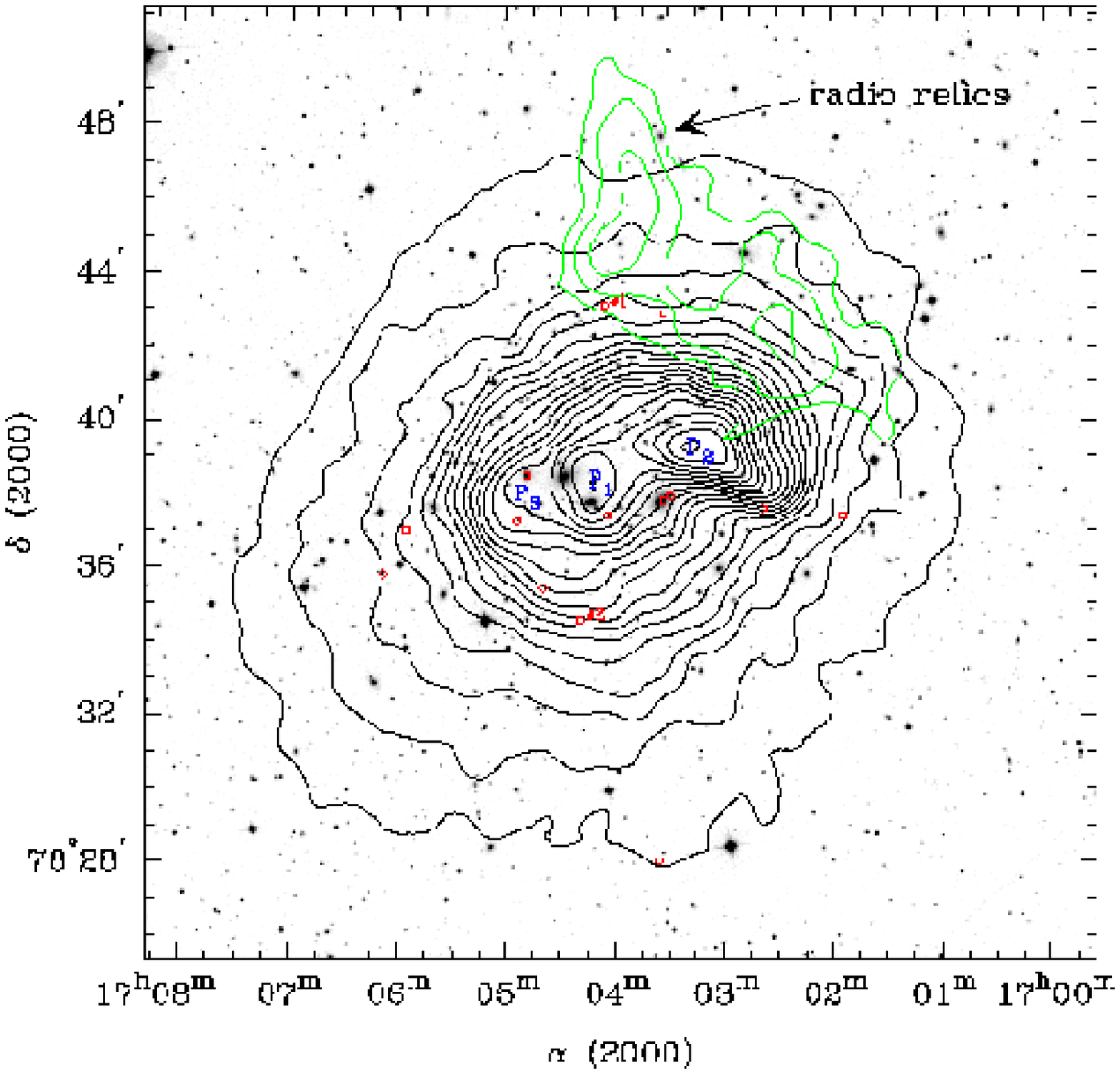}
\vspace{0.5cm}
  \caption{The 0.5 - 7 keV ACIS-I/S co-added contours (point sources excluded)
overlaid on a Digitized Sky Survey second generation red (DSS II hereafter)
image. The ACIS-I/S image was smoothed by a variable-width Gaussian whose
$\sigma$ varies from 20$''$ at the peak to 30$''$ near the edges of the image.
The contour levels are linear from 0.03 to 0.625 c/ks/pixel (pixel size:
$3.934''$). The outermost contour is affected by the edges of the CCD chips.
P$_{1}$, P$_{2}$ and P$_{3}$ correspond to the peak region of the main cluster,
the peak region of the subcluster and the "shoulder" respectively.
The small red circles represent the bright point sources ($>$ 20 counts)
detected by ACIS-I/S, including the three corresponding to member
galaxies. \#1 and \#2 are the two brightest point sources
(see $\S$3.1). The green contours show the position of the strongest radio
relic in this source (from NVSS 20 cm survey).
There are more diffuse radio relics around P$_{2}$ and P$_{3}$ (Bridle \&
Fomalont 1976; R94).
    \label{fig1}}
\end{figure}

\clearpage

\begin{figure}
\vspace{-1cm}
\hspace{-2cm}
  \includegraphics[height=1.2\linewidth,angle=270]{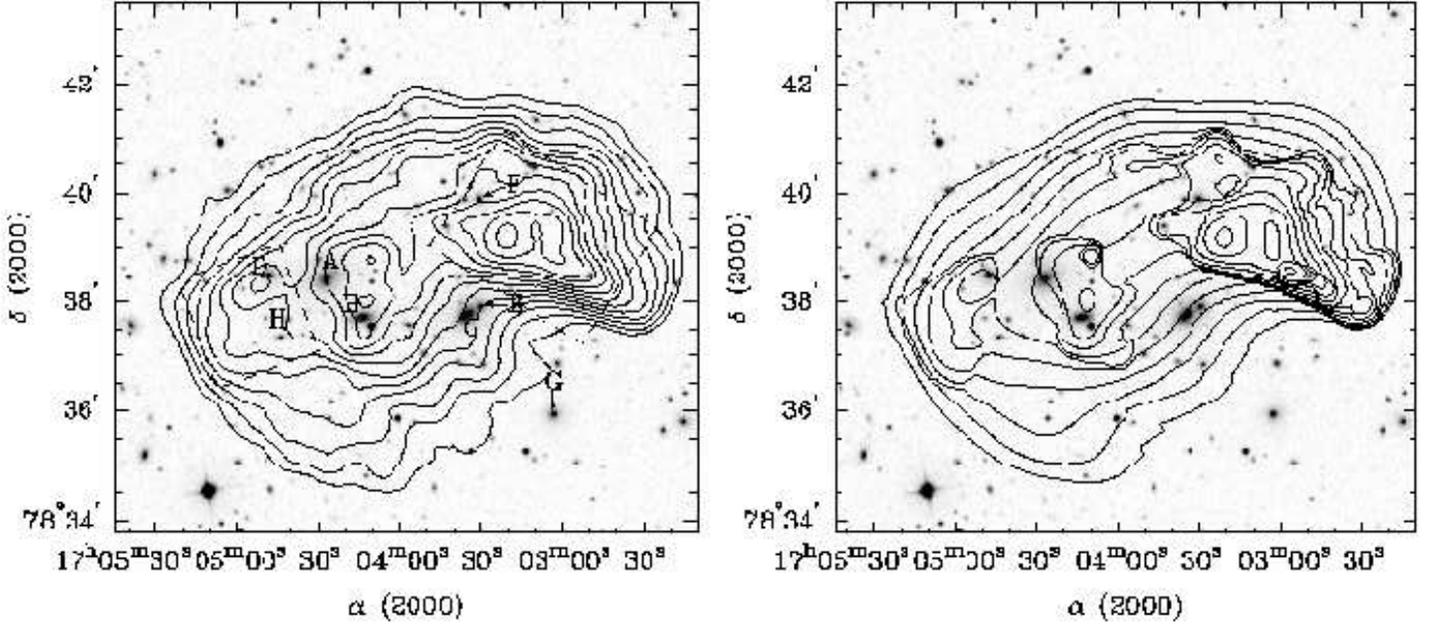}
\vspace{-1.5cm}
  \caption{The left one: central region of the 0.5 - 7 keV ACIS-I/S co-added
image (linear-spaced contours) overlaid on the DSS II image. The ACIS-I/S
image was smoothed by a $15''$ $\sigma$ Gaussian. We used A - H to label the
big galaxies around the center. Their redshifts are 0.0594, 0.0564, 0.0586,
0.0643, 0.0587, 0.0586, 0.0553 and 0.0508 from A to H respectively (from NASA
extragalactic database --- NED). The arrows for E, F and G (all radio
head-tail sources) show the rough directions of the head-tails. Notice the
"shoulder" east of A, the southern "edge" of P$_{2}$ (see Fig. 1) and
the SW protrusion from P$_{2}$. The dash boxes represent the two
substructures around the center and analysis was made to these regions
in section $\S$3.6. The right one: the reconstructed image (linear-spaced contours) by wavelet
analysis in the same region as the left. This image also reveals the
complicated structure at the center with very similar pattern as the
left. The surface density discontinuity is enhanced in this wavelet
reconstructed image.
    \label{fig2}}
\end{figure}

\clearpage

\begin{figure}
\vspace{-2.3cm}
\hspace{-1cm}
  \includegraphics[height=1.2\linewidth,angle=270]{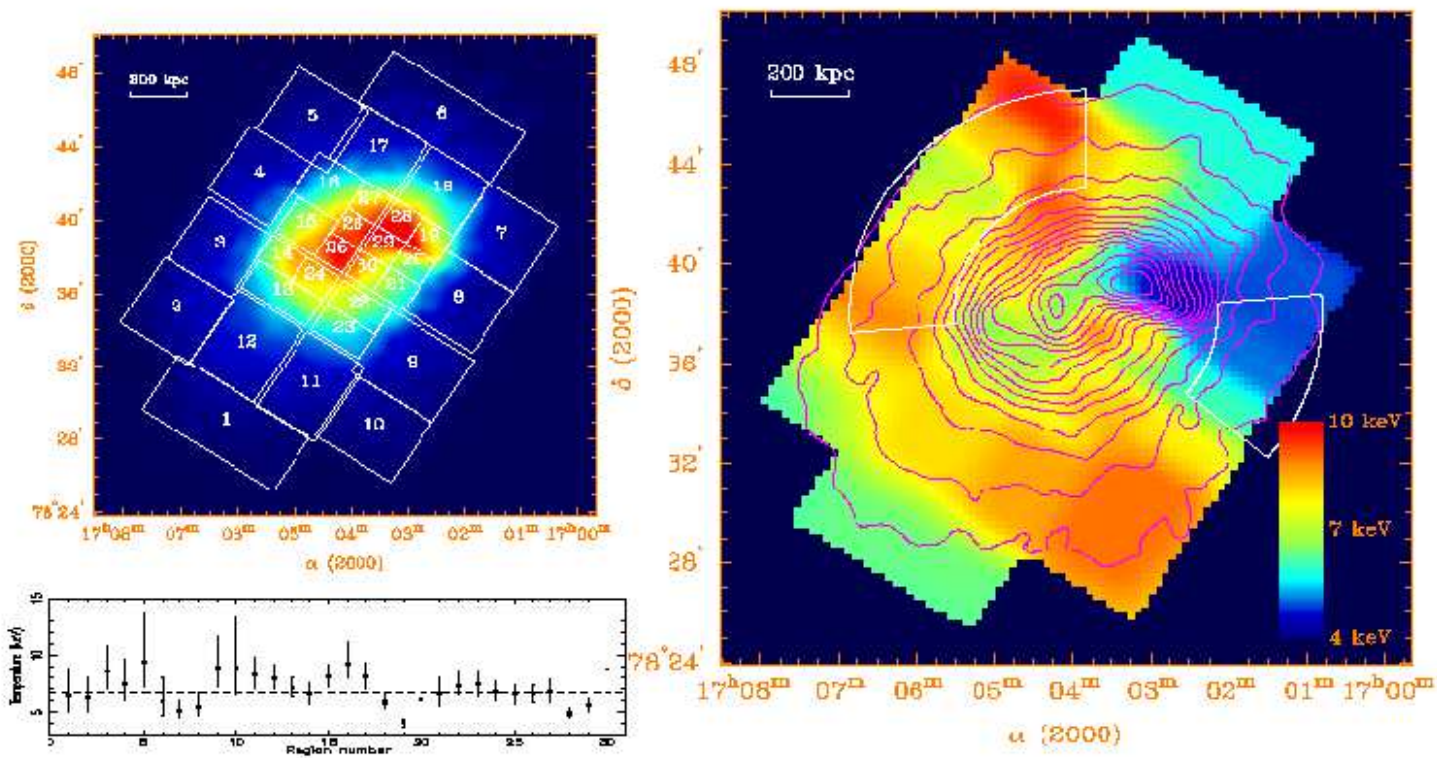}
\vspace{-2.5cm}
  \caption{Temperature map. Upper left panel: the
first set of regions we chose (the second and third sets are regions
shifted high region size relative to the first set); Lower left panel:
the fitting results for the first (of three) set of regions (The results for the
second and third sets agree with those of the first set); the dash line
represents the average temperature measured in the central 8.3$'$ square
ACIS-S3 FOV --- 6.7 keV. Right panel: The
adaptively smoothed image of the final temperature map. The average smoothing
scale ($\sigma$) is 0.5$'$. This map
is the average of the results of three overlapping sets of
region described in the text, not simply the smoothing from the results
shown on the left. The errors on this temperature map can be estimated from
the lower left panel. Two sectors shown in white are the
"hot spots" reported by Briel \& Henry (1994). \label{fig3}}
\end{figure}

\clearpage

\begin{figure}
\vspace{-2.2cm}
\hspace{-4.5cm}
  \includegraphics[angle=270]{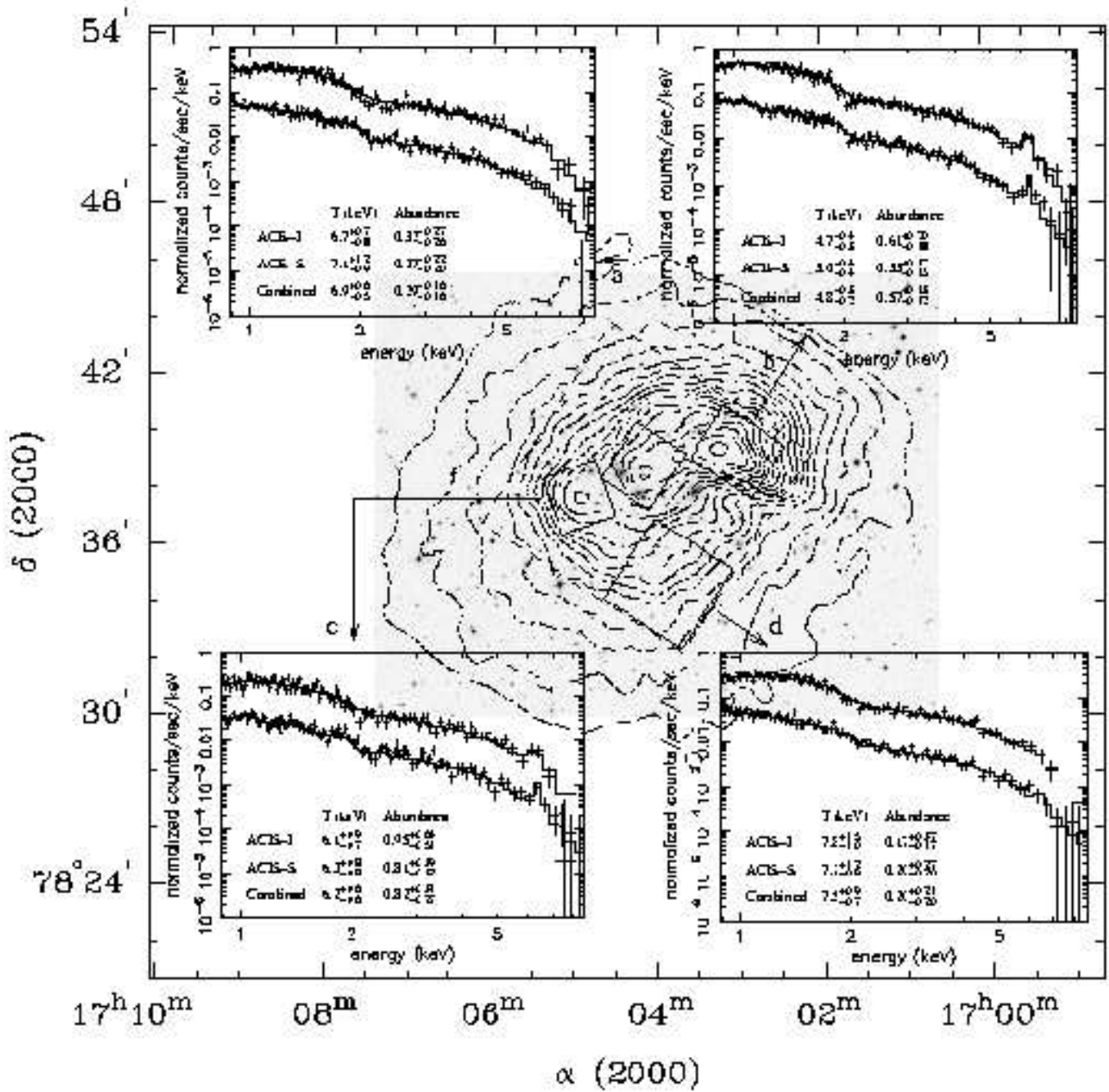}
\epsscale{0.8}
\vspace{-10mm}
  \caption{Iron abundance in several interesting regions: a - P$_{1}$; b - P$_{2}$;
c - P$_{3}$; d - a southern region.
Region c is same as what we used in Fig. 2 for P$_{3}$. Regions a and b cover
basically same regions as what we used in Fig. 2 for P$_{1}$ and P$_{2}$
but are not crossing the chips.
In each small box, the upper spectrum is that of ACIS-I and the lower one
is that of ACIS-S rescaled by 0.1.
Both spectra and fitting results suggest that P$_{2}$ and P$_{3}$
have more iron than P$_{1}$
and the southern region d. The square dash region is the FOV of
ACIS-S3 chip in observation 965. The contours are from ACIS-I observation
1386 and the gray image is DSS II.
    \label{fig4}}
\end{figure}

\clearpage

\begin{figure}
  \includegraphics[height=0.95\linewidth,angle=270]{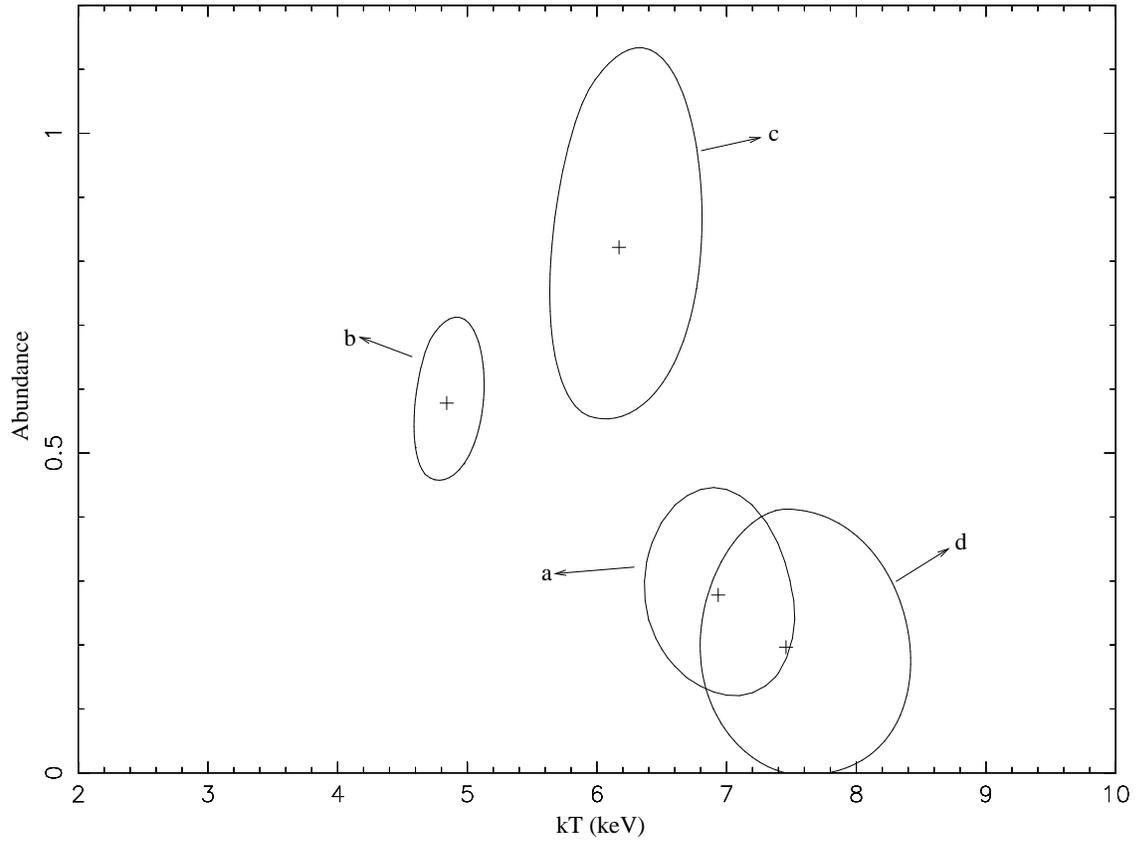}
\vspace{0.5cm}
  \caption{The joint abundance-temperature 90\% confidence regions for
four regions in Fig. 4.
    \label{fig5}}
\end{figure}

\clearpage

\begin{figure}
\vspace{-1cm}
\hspace{-2.7cm}
  \includegraphics[height=1.05\linewidth,angle=270]{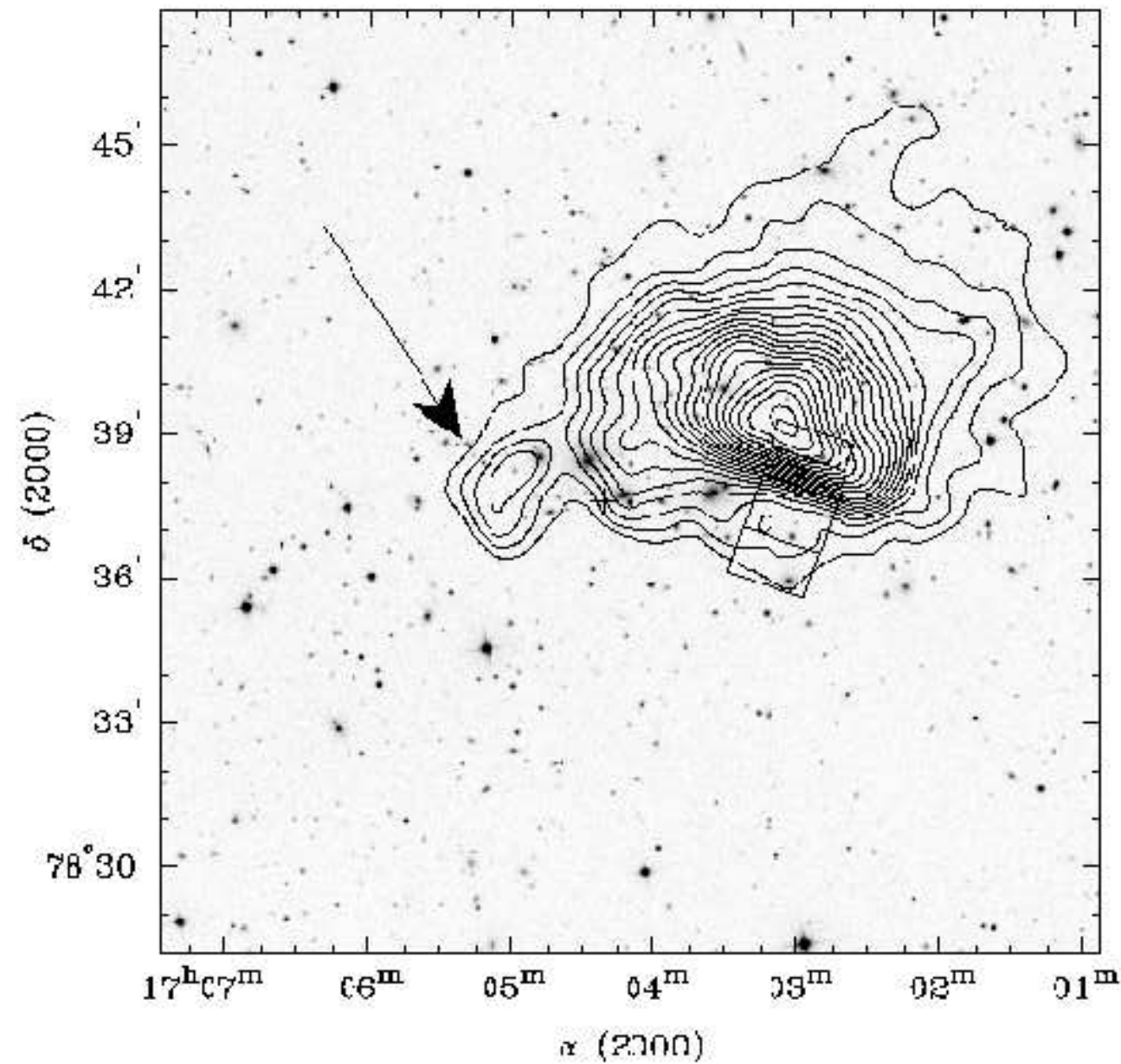}
  \caption{Residual after removing the main cluster represented by the
best-fit $\beta$-model. The residual was smoothed by
a variable-width Gaussian whose $\sigma$ varies from 16$''$ at the peak
to 30$''$ near the edges of the image. The contour levels are linearly from
0.03 to 0.98 of the maximum with an interval 0.05 of the maximum. The
"shoulder" is very prominent --- pointed by the arrow.
The edge at the south of the subcluster peak is very clear. The
four regions in line are those that we made temperature measurement
in $\S$3.6. The cross is the derived geometrical center of the main
cluster.
    \label{fig6}}
\end{figure}

\clearpage

\begin{figure}
  \plotone{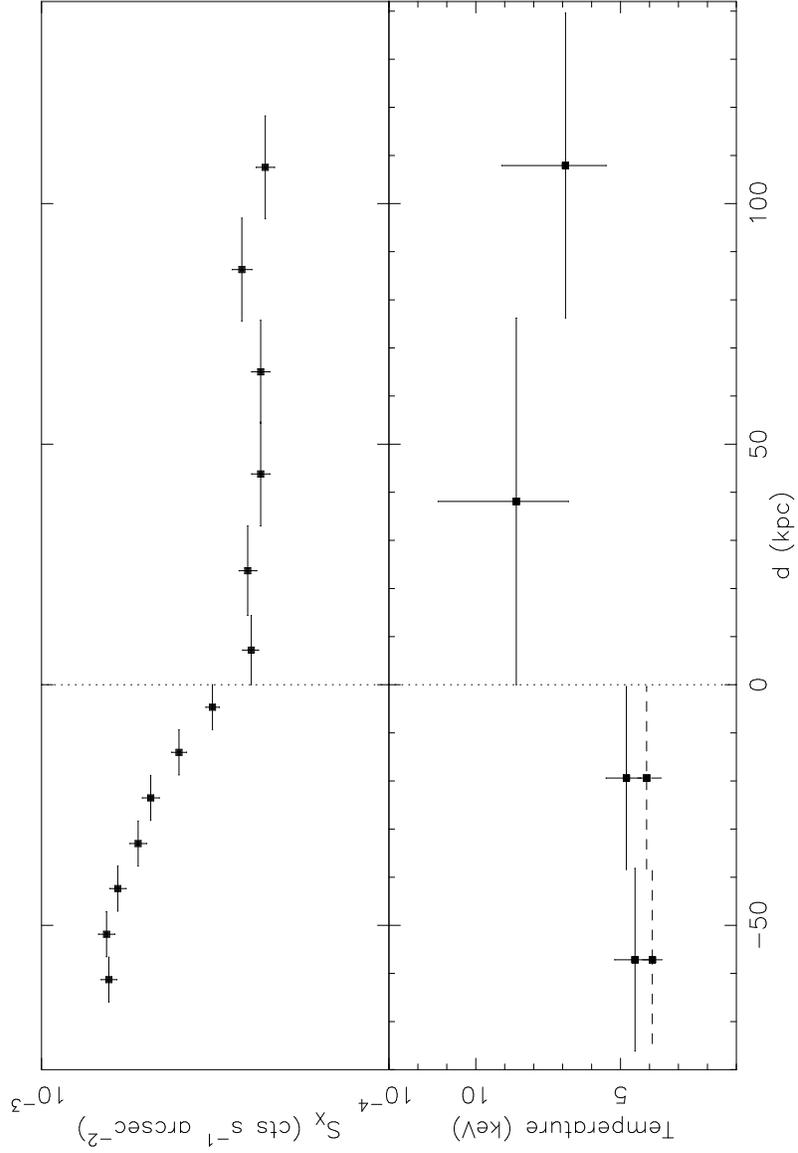}
  \caption{The projected surface brightness profile (measured in co-add
ACIS-I/S image) and temperature across the edge. The regions are shown
in Fig. 6. The distance is measured relative to
the edge. The temperature fits represented by the dash lines are
the decomposed temperature using the method mentioned in $\S$3.5.
    \label{fig7}}
\end{figure}

\clearpage

{\footnotesize
\renewcommand{\arraystretch}{1.4}
\renewcommand{\tabcolsep}{1mm}
\begin{center}
TABLE 1
\vspace{1mm}

{\sc Observations}
\vspace{1mm}

\begin{tabular}{c|c|c|c}
\hline \hline
 Obs. ID & 1386 & 965 & 1521\\\hline
 Obs. Mode & ACIS-I, & ACIS-S, & ACIS-S, \\
           & 012378 & 236789 & 136789 \\
 Observing time & Oct. 13, 99& Oct. 13, 99& Feb. 27, 00\\
 Total exposure (s) & 15809 & 15717 & 2976  \\
 Effective exposure (s) & 9966 & 8256(S3), & - \\
                      &  &  9458(S2,S4) & \\
 RA (2000) & 17:04:14 & 17:04:36 & 17:04:22 \\
 DEC (2000) & 78:37:52 & 78:36:18 & 78:39:59 \\
\hline\hline
\end{tabular}
\end{center}}

{\footnotesize
\renewcommand{\arraystretch}{1.4}
\renewcommand{\tabcolsep}{1mm}
\begin{center}
TABLE 2
\vspace{1mm}

{\sc Decomposition of P$_{2}$ and P$_{3}$}
\vspace{1mm}

\begin{tabular}{c|c|c|c|c|c|c|c|c}
\hline \hline
 Region & T$^{\rm a}$ & Abund.$^{\rm a}$ & T$^{\rm b}$ &Abund.$^{\rm b}$ &
 T$^{\rm c}$ & Abund.$^{\rm c}$ & T$^{\rm d}$ & Abund.$^{\rm d}$\\
   & (keV) &    &  (keV) &  & (keV) &  & (keV) &\\\hline
 P$_{2}$ & 5.0$^{+0.3}_{-0.3}$ & 0.51$^{+0.12}_{-0.11}$ & 4.3$^{+0.3}_{-0.3}$ &
 0.59$^{+0.17}_{-0.14}$ & 4.2$^{+0.3}_{-0.3}$ &
0.56$^{+0.16}_{-0.14}$ & 4.0$^{+0.3}_{-0.3}$ & 0.55$^{+0.17}_{-0.13}$ \\
 P$_{3}$ & 6.2$^{+0.6}_{-0.6}$ & 0.82$^{+0.31}_{-0.27}$ & 5.0$^{+1.3}_{-1.1}$ &
 $>$ 0.75 & 4.5$^{+1.0}_{-1.0}$ &$
>$ 0.68 & 3.9$^{+1.1}_{-0.8}$ &$>$ 0.55 \\\hline\hline
\end{tabular}
\begin{flushleft}
\leftskip 27pt
$^{\rm a}$ Uncorrected\\
$^{\rm b}$ After the decomposition (7 keV assumption, see text)\\
$^{\rm c}$ After the decomposition (8 keV assumption, see text)\\
$^{\rm d}$ After the decomposition (9 keV assumption, see text)\\
\end{flushleft}
\end{center}}


\begin{references}

 \reference{} Anders E., \& Grevesse N. 1989, Geochimica et Cosmochimica Acta, 53, 197

 \reference{} Arimoto, N., \& Yoshii, Y. 1987, A\&A, 173, 23

 \reference{} Bekki, K. 1999, ApJ, 510, 15

 \reference{} Bliton, M., Rizza, E., Burns, J. O., Owen, F. N., \& Ledlow, M. J. 1998, MNRAS, 301, 609

 \reference{} Bridle, A., \& Fomalont, E. 1976, A\&A, 52, 107

 \reference{} Bridle, A., Fomalont, E., Miley, G., \& Valentijn, E. 1979, A\&A, 80, 201

 \reference{} Briel, U. G. et al. A\&A, 1991, 246, L10

 \reference{} Briel, U. G., \& Henry, J. P. 1994, Nature, 372, 439

 \reference{} De Grandi, S., \& Molendi, S. 1999, ApJ, 527, L25

 \reference{} Ezawa, H. et al. 1997, ApJ, 490, L33

 \reference{} Fabian, A. C., \& Daines, S. J. 1991, MNRAS, 252, 17P

 \reference{} Fabricant, D. G., \& Kent, S. M., \& Kurtz, M. J. 1989, ApJ, 336, 77

 \reference{} Fukazawa, Y. et al. 1994, PASJ, 46, L55

 \reference{} Fukazawa, Y. et al. 1998, PASJ, 50, 187

 \reference{} Fusco-Femiano, R. et al. 2000, ApJ, 534, L7

 \reference{} Giacconi, R. et al. 2001, ApJ submitted (astro-ph/0103014)

 \reference{} Gonz\'alez-Casado, G., Mamon, G. A., \& Salvador-Sol\'e, E. 1994, ApJ, 433, L61  

 \reference{} Kitayama, T., Sasaki, S., \& Suto, Y. 1998, PASJ, 50, 1

 \reference{} Markevitch, M. 1996, ApJ, 465, L1 (M96)

 \reference{} Markevitch, M., \& Vikhlinin, A. 1997a, ApJ, 474, 84

 \reference{} Markevitch, M., \& Vikhlinin, A. 1997b, ApJ, 491, 467 

 \reference{} Markevitch, M. et al. 2000, ApJ, 541, 542

 \reference{} Markevitch, M. 2001, \chandra\ calibration memo, http://asc.harvard.edu/cal/ , "ACIS", "ACIS background"

 \reference{} Markevitch, M., \& Vikhlinin, A. 2001, ApJ, in press  (astro-ph/0105093)

 \reference{} Matsumoto, H. et al. 1996, PASJ, 48, 201

 \reference{} Mazzotta, P., Markevitch, M., Vikhlinin, A., Forman, W. R., David, L. P., \& Vanspeybroeck, L. 2001, ApJ, 555, 205

 \reference{} Miyaji et al. 1993, ApJ, 419, 66

 \reference{} Molendi, S., De Grandi, S., \& Fusco-Femiano, R. 2000, ApJ, 533, L43 (MDF00)

 \reference{} Owen, F., N., Ledlow, M., J., Keel, W., C., \& Morrison, G., E. 1999, AJ, 118, 633

 \reference{} Roettiger, K., Burns, J. O., \& Pinkney, J. 1995, ApJ, 453, 634

 \reference{} Roettiger, K., Loken, C., \& Burns, J. O. 1997, ApJS, 109, 307

 \reference{} Roettiger, K., Stone, J. M., \& Burns, J. O. 1999,ApJ, 518, 594

 \reference{} R$\ddot{\rm o}$ttgering, H., Snellen, I., Miley, G., De Jong, J. P., Hanisch, R. J., \& Perley, R. 1994, ApJ, 436, 654 (R94)

 \reference{} Slezak, E., Durret, F., \& Gerbal, D. 1994, AJ, 108, 1996

 \reference{} Takizawa, M. 1999, ApJ, 520, 514

 \reference{} Takizawa, M. 2000, ApJ, 532, 183

 \reference{} Tribble, P. C. 1993, MNRAS, 263, 31

 \reference{} Trimble, V. 2000, in Allen's Astrophysical Quantities, fourth edition, ed. A. N. Cox (Springer AIP press), p577

 \reference{} Vikhlinin, A., Forman, W., Jones, C. 1997, 474, L7

 \reference{} Vikhlinin, A. 2000, \chandra\ calibration memo, http://asc.harvard.edu/cal/Links/Acis/acis/Cal\_prods/qe/12\_01\_00/

 \reference{} Vikhlinin, A., Markevitch, M., \& Murray, S. S. 2001a, ApJ, 551, 160

 \reference{} Vikhlinin, A., Markevitch, M., Forman, W., \& Jones, C. 2001b, ApJL, 555, 87

 \reference{} White, D. A. 2000, MNRAS, 312, 663

 \reference{} Xu, H. et al. 1997, PASJ, 49, 9
\end{references}
\end{document}